\documentclass[showpacs,amsmath,amssymb,twocolumn,pra,superscriptaddress]{revtex4-1}
\usepackage{amssymb}

\usepackage[dvips]{graphicx} 
\usepackage{enumerate}
\usepackage{epsfig}
\usepackage{subfigure}
\usepackage{xcolor}
\usepackage[T1]{fontenc}
\usepackage[expert]{mathdesign}
\usepackage{fullpage}
\usepackage{amsthm,amsfonts,amssymb,amscd,mathrsfs,eufrak,xspace,framed}
\usepackage{amsmath}
\usepackage{color}
\usepackage{setspace}
\usepackage{url}
\usepackage{enumitem}
\usepackage{slashbox}

\begin{document}
\title{Practical gigahertz quantum key distribution robust against channel disturbance}

\author{Shuang Wang}

\author{Wei Chen}
\email{weich@ustc.edu.cn}

\author{Zhen-Qiang Yin}
\email{yinzq@ustc.edu.cn}

\author{De-Yong He}
\affiliation{Key Laboratory of Quantum Information, CAS, University of Science and Technology of China, Hefei 230026, China}
\affiliation{Synergetic Innovation Center of Quantum Information and Quantum Physics, University of Science and Technology of China, Hefei 230026, China}
\affiliation{State Key Laboratory of Cryptology, P. O. Box 5159, Beijing 100878, China}

\author{Cong Hui}
\affiliation{Department of Modern Physics, University of Science and Technology of China, Hefei 230026, China}

\author{Peng-Lei Hao}
\affiliation{Anhui Asky Quantum Technology Co. Ltd., Wuhu 241002, China}

\author{Guan-Jie Fan-Yuan}

\author{Chao Wang}

\author{Li-Jun Zhang}
\affiliation{Key Laboratory of Quantum Information, CAS, University of Science and Technology of China, Hefei 230026, China}
\affiliation{Synergetic Innovation Center of Quantum Information and Quantum Physics, University of Science and Technology of China, Hefei 230026, China}
\affiliation{State Key Laboratory of Cryptology, P. O. Box 5159, Beijing 100878, China}

\author{Jie Kuang}
\affiliation{Department of Modern Physics, University of Science and Technology of China, Hefei 230026, China}

\author{Shu-Feng Liu}
\affiliation{Anhui Asky Quantum Technology Co. Ltd., Wuhu 241002, China}

\author{Zheng Zhou}
\affiliation{Key Laboratory of Quantum Information, CAS, University of Science and Technology of China, Hefei 230026, China}
\affiliation{Synergetic Innovation Center of Quantum Information and Quantum Physics, University of Science and Technology of China, Hefei 230026, China}
\affiliation{State Key Laboratory of Cryptology, P. O. Box 5159, Beijing 100878, China}

\author{Yong-Gang Wang}
\affiliation{Synergetic Innovation Center of Quantum Information and Quantum Physics, University of Science and Technology of China, Hefei 230026, China}
\affiliation{Department of Modern Physics, University of Science and Technology of China, Hefei 230026, China}

\author{Guang-Can Guo}

\author{Zheng-Fu Han}
\affiliation{Key Laboratory of Quantum Information, CAS, University of Science and Technology of China, Hefei 230026, China}
\affiliation{Synergetic Innovation Center of Quantum Information and Quantum Physics, University of Science and Technology of China, Hefei 230026, China}
\affiliation{State Key Laboratory of Cryptology, P. O. Box 5159, Beijing 100878, China}



\begin{abstract}
	Quantum key distribution (QKD) provides an attractive solution for secure communication. However, channel disturbance severely limits its application when a QKD system is transfered from the laboratory to the field. Here, a high-speed Faraday-Sagnac-Michelson QKD system is proposed that can automatically compensate for the channel polarization disturbance, which largely avoids the intermittency limitations of environment mutation. Over a 50-km fiber channel with 30-Hz polarization scrambling, the practicality of this phase-coding QKD system was characterized with an interference fringe visibility of $99.35\%$ over 24 hours, and a stable secure key rate of 306k  bits/s over 7 days without active polarization alignment.
\end{abstract}

\maketitle

Quantum key distribution (QKD) allows two remote parties (named Alice and Bob) to share secure keys in the presence of an eavesdropper. Its unconditional security is guaranteed by the fundamental laws of quantum physics. As the first commercial application of quantum physics at the single-quantum level \cite{gisin2002}, the practicality of QKD becomes important when we transfer a QKD system from the laboratory to the field. Contrary to laboratory conditions, field environments are complex and volatile, which would continually interrupt or even terminate the operation of QKD \cite{xu2009,SECOQC09,ol2010,swiss,tokyo,hcw,yuan2012,yuan2015}, and would severely limit its application.

In QKD systems deployed over telecom fiber networks, the field environments typically break down into two parts: operating conditions of QKD units, and environments of installed fiber channels that are used to transmit quantum states. By careful design, the operating conditions of QKD units can be well controlled, i.e., isolated from complex environments. While, the environments of installed fiber channels are uncontrollable and ever-changing, quantum states are inevitably disturbed during transmission in fiber channels. More crucially, channel disturbance induced by the external environment accumulates with distance \cite{ding}. Due to volatile field environments, variations in optical path length (or time delay) and fiber channel birefringence \cite{hcw,yuan2012,yuan2015,ding} are two obstacles preventing long-term continuous operation of QKD systems. However, optical path length variations in fiber channels are relatively slow and could be well tracked \cite{yuan2012} or compensated \cite{hcw} with simple synchronization schemes. Thus, birefringence variations in fiber channels become the most challenging channel disturbance in practical QKD systems.

Channel disturbance countermeasures in QKD systems can be classified into active and passive categories. In QKD systems employing active countermeasures, polarization basis alignment modules are added to compensate for birefringence variations in fiber channels \cite{yuan2012,yuan2015,ding} and are essential for low quantum bit error rate (QBER). In QKD systems adopting passive countermeasures, birefringence variations in fiber channels can be automatically compensated, which largely avoid the intermittency limitations of environment mutation. The “Plug-and-Play” system \cite{PP} and Faraday-Michelson system \cite{fm} are typical ones, and they feature low QBER and excellent long-term stability \cite{hcw,swiss}. These QKD systems have been deployed in complex field fiber networks and show additional advantages, such as fast establishment of new QKD links \cite{hcw}. However, a system's operation speed is limited by its structure. Typical operation speeds of a two-way “Plug-and-Play” system and a one-way Faraday-Michelson system are only 5 MHz \cite{SECOQC09} and 200 MHz, respectively.

Here, we overcome this speed limitation by improving the structure of a Faraday-Michelson QKD system. The newly developed QKD system is based on the asymmetric Faraday-Sagnac-Michelson interferometer structure. It is intrinsically stable against channel disturbance and has high-speed support. In a 50-km single mode fiber channel with polarization scrambling, the interference fringe visibility was tested over 24 hours, and gigahertz quantum key distribution was implemented over 7 days without active polarization alignment. The practicality of this gigahertz QKD system is characterized by the stabilities of visibility, QBERs, and overall gains under channel disturbance. 

\begin{figure}[tbp] 
	\includegraphics[width=1.0\columnwidth]{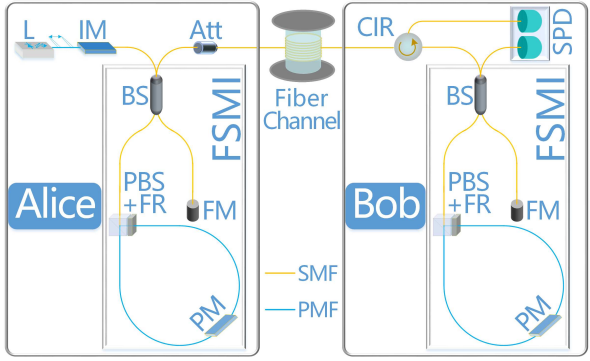}\\
	\caption{Schematic of Faraday-Sagnac-Michelson QKD system.}
	\label{fig:1}
\end{figure}

The Faraday-Sagnac-Michelson QKD system is outlined in Fig.~\ref{fig:1}. At Alice's site, optical pulses with random phases are generated by a semiconductor gain-switched laser (L) \cite{gs1,gs2}. The three intensity levels required for the decoy-state method \cite{decoy0,decoy1,decoy2} are created using an intensity modulator (IM). Photon phase-encoding is performed by controlling the phase modulator (PM) in her asymmetric Faraday-Sagnac-Michelson interferometer (FSMI). The photon pulses are attenuated to the single-photon level by an attenuator (Att). These coded weak pulses are then transmitted to Bob via a single mode fiber channel. At Bob's site, photon phase-decoding is achieved by controlling the PM in his FSMI. Through a three-port circulator (CIR), photons from two outputs of Bob's interferometer are detected by a double-channel single-photon detector (SPD).

\begin{figure}[tbp]
	\includegraphics[width=1.0\columnwidth]{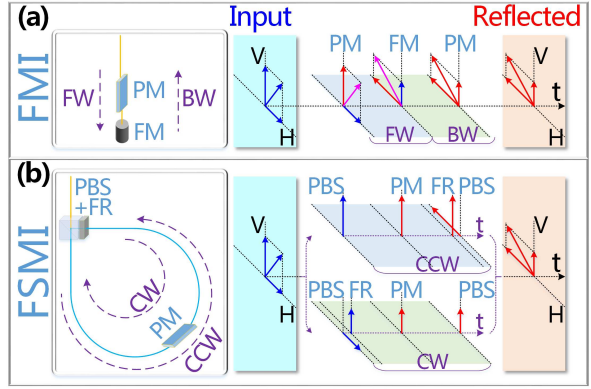}\\
	\caption{Configuration of the long arm and phase modulation process in (a) FMI and (b) FSMI. H and V represent horizontal and vertical polarizations, respectively.}
	\label{fig:2}
\end{figure}

In comparison with our previous Faraday-Michelson QKD system \cite{fm}, two asymmetric Faraday-Michelson interferometers (FMI) were replaced by two FSMIs in the newly developed system. The structure of the FMI and FSMI is similar: only one $50/50$ beam splitter (BS) was used and the optical pulse passes bidirectionally in two arms, which are called the short arm and long arm, respectively. The short arms in the FMI and FSMI are the same. The short arm contains just one Faraday mirror (FM), which is composed of a $45^{\circ}$ Faraday rotator (FR) and a total reflection mirror. The only difference lies in the long arm (see the left panel in Fig.~\ref{fig:2}). For FMI, the long arm is composed of a PM and a FM. For FSMI, the long arm has a Sagnac configuration, which is composed of a polarization beam splitter (PBS) with a $90^{\circ}$ FR and a PM. In Fig.~\ref{fig:1} and Fig.~\ref{fig:2}, the label ‘PBS+FR’ indicates that the PBS and the $90^{\circ}$ FR are packaged together as one optical component. For the Sagnac configuration in the FSMI, the input (also output) fiber is a single mode fiber (SMF), while the fiber in it is a polarization maintaining fiber (PMF). 

The phase modulation effects in the FMI and FSMI are equivalent if all optical components are perfect. To demonstrate this, we only need to verify the equivalence of input and reflected pulses of the long arm between FMI and FSMI, since the configuration of the long arm is the only difference between FMI and FSMI, and PM lies in the long arm. As shown in the right panel of Fig.~\ref{fig:2}, for a given input pulse with arbitrary polarization, the reflected pulses in the long arm of the FMI and FSMI are identical in terms of their added phase and polarization direction. For the convenience of description, PM is assumed to be effective only in the vertical direction and the added phase is denoted as $\phi$. The phase modulation process in the long arm of the FMI is shown in the right portion of Fig.~\ref{fig:2}(a). When the input pulse passes along the forward (FW) direction, a phase $\phi$ is added on its vertical component by PM. After being reflected by the FM, the original horizontal component is rotated into the vertical direction and a phase $\phi$ is added while the pulse passes along the backward (BW) direction. Compared with the input pulse, a phase $\phi$ is added and the polarization direction of the reflected pulse is rotated $90^{\circ}$ in the long arm of the FMI. The phase modulation process in the long arm of the FSMI is shown in the right portion of Fig.~\ref{fig:2}(b). After entering the Sagnac configuration, the input pulse is separated into two parts by the PBS. The vertical part propagates along the counter clockwise (CCW) direction, is first added a phase $\phi$ by PM and then rotated into the horizontal direction by the  $90^{\circ}$ FR. Meanwhile, the horizontal part propagates along the clockwise (CW) direction, is first rotated into the vertical direction by the $90^{\circ}$ FR and then added a phase $\phi$ by PM. Finally, these two parts are combined as a reflected pulse by the same PBS. Therefore, compared with the input pulse, a phase $\phi$ is added and the polarization direction of reflected pulse is also rotated $90^{\circ}$ in the long arm of the FSMI, just the same as the long arm of the FMI.

\begin{figure}[tbp]
	\includegraphics[width=1\columnwidth]{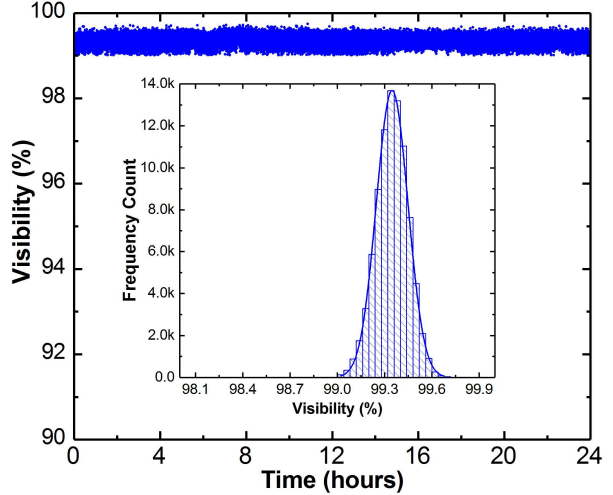}\\
	\caption{Measured visibility over 24 hours. The inset is the histogram of the visibility distribution.}
	\label{fig:3}
\end{figure}

We can conclude that the phase-coding QKD system based on the asymmetric FSMI is also robust against channel disturbance, just like the system based on the asymmetric FMI, whose intrinsic stability has been demonstrated in theory \cite{fm} and in practice \cite{xu2009,ol2010,hcw,ding}. To experimentally demonstrate the stability of a Faraday-Sagnac-Michelson QKD system, we measured its interference fringe visibility with polarization scrambling in the fiber channel. The driving voltage of Bob's PM scans from $0$ V to $9$ V with a step of about $0.01$ V. At each step, the count in the first channel of the SPD is recorded after waiting for $2^{19}$ clock cycles. Once the driving voltage scan procedure is complete, the corresponding interference fringe visibility is calculated using
\begin{equation}
\label{eq:1}
V = \frac{C_{max}-C_{min}}{C_{max}+C_{min}},
\end{equation}
where $C_{max}$ and $C_{min}$ are the maximum and minimum counts in one channel of the SPD, respectively. The driving voltage scan procedure is performed approximately once a second. The frequency of polarization scrambling is $30$ Hz. Fig.~\ref{fig:3} shows the measured visibility over 24 hours. The Faraday-Sagnac-Michelson QKD system achieves a high visibility of $99.35\%$ on average. The $0.65\%$ deviation from perfect visibility primarily originates from apparatus imperfections, such as interferometer misalignment, finite precision of the scanned driving voltage, and dark counts of the SPD \cite{yuan2008}. The visibility of interference fringes is very stable. As shown in the inset of Fig.~\ref{fig:3}, the visibility distribution is concentrated around $99.35\%$ with a standard deviation of $0.12\%$.

In addition to the intrinsic stability, the QKD system based on the asymmetric FSMI also has the following three advantages compared with the previous one based on the asymmetric FMI. \textit{(I) High-speed support.} Speed limitation is the toughest challenge we encountered with the FMI. Since the long arm of the FMI has a forward and backward configuration, the corresponding QKD system has difficulty supporting gigahertz clock rates using only a simple phase modulation scheme, in which PM driving voltage remains unchanged during each clock period. Considering the rise and fall time of the driving voltage, the minimum clock rate of the QKD system is about twice that of the flat phase modulation duration, which depends on the lengths of the PM and FM, and the fiber length between them. Even if the PM could be directly placed adjacent to the FM, the maximum supported speed is only 500 MHz. While for the long arm of the FSMI, PM is placed in the middle of the Sagnac configuration, two parts of the input pulse arrived at the PM simultaneously and with the same polarization direction. Thus, the QKD system based on the asymmetric FSMI can support gigahertz clock rates by employing a simple phase modulation scheme. \textit{(II) Low insertion loss.} The insertion loss is very important in Bob's asymmetric interferometer, and its primary source comes from the PM. In the FMI, two polarization parts of the input pulse pass through the PM twice (once in FW direction and once in BW direction). While in the FSMI, two polarization parts of the input pulse become the same polarization when they pass through the PM. In terms of insertion loss, this is equivalent to the input pulse passing through the PM once.\textit{(III) Lower requirements for PM.} In the FMI, both orthogonal polarization components of input light should be able to pass through the PM. The only requirement for the PM in the FSMI structure is that both input and output fibers are PMF, which is consistent with conventional PM requirements.

\begin{figure}[tbp]
	\includegraphics[width=1.0\columnwidth]{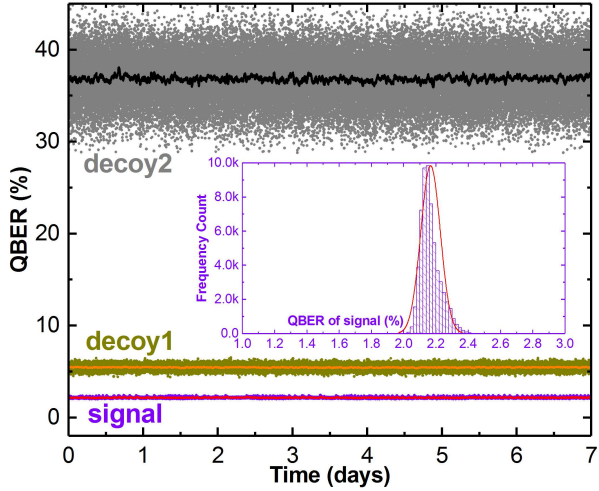}\\
	\caption{QBERs of the signal and two different decoy pulses. The inset is a QBER frequency histogram for the signal pulses.}
	\label{fig:4}
\end{figure}

We perform a gigahertz QKD experiment over a 50-km fiber channel with polarization scrambling, by employing a phase-coding system with the FSMI. BB84 protocol \cite{bb84} is implemented with the three intensity decoy-state method \cite{ma2005}, in which signal pulses of 0.48 photons per pulse are sent with 29/32 probability, and two different decoy pulses (denoted as decoy1 and decoy2) of 0.07 and 0.002 photons per pulse are sent with 2/32 and 1/32 probability, respectively. The double-channel SPD is based on InGaAs/InP avalanche photodiodes and is operated in gated Geiger mode with the sine-wave filtering method \cite{lpf}. To achieve low afterpulse probability (or low error rate \cite{yuan2008}), we add a width discriminator to remove the wider filtered avalanche signals in each channel \cite{spd}. The average detection efficiency is greater than $20\%$, the afterpulse probability is less than $1.1\%$, and the total dark count rate is less than $2\times10^{-6}$ per gate.
 
QBERs of the signal and two different decoy pulses obtained over 7 days operation are shown in Fig.~\ref{fig:4}. Each point stands for the measured value approximately every 10 seconds. For the signal, decoy1, and decoy2 pulses, the distributed regions of these points are different, which primarily originates from the statistical fluctuations \cite{yuan2010}. The solid lines are obtained by averaging every 200 adjacent points and clearly show QBER stability under channel disturbance. The mean QBERs for the signal, decoy1, and decoy2 pulses are $2.17\%$, $5.46\%$, and $36.83\%$, respectively. This low QBER for the signal pulses demonstrates the effectiveness of the newly developed system at 1 GHz, while the QBER for the signal pulses was over $3\%$ for our previous Faraday-Michelson system at 200 MHz. As shown in the inset of Fig.~\ref{fig:4}, the QBER distribution of signal pulses is concentrated around its mean value (between $2.0\%$ and $2.5\%$). Since the total dark count rate is relatively low, error counts for the decoy1 and decoy2 pulses primarily originate from afterpulses. These results agree well with the stability of interference fringe visibility.

\begin{figure}[tbp]
	\includegraphics[width=1.0\columnwidth]{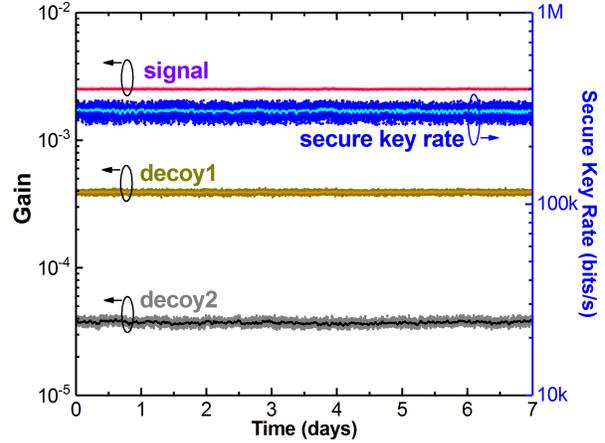}\\
	\caption{The overall gain (left axis) of the signal and two different decoy pulses respectively, and secure key rate (right axis).}
	\label{fig:5}
\end{figure}

The corresponding overall gains \cite{ma2005} of the signal, decoy1 and decoy2 pulses, and a secure key rate are shown in Fig.~\ref{fig:5}. Just as in Fig.~\ref{fig:4}, measurements for each point were collected over approximately 10 seconds, and the solid lines are obtained by averaging every 200 adjacent points. The gain stability under channel disturbance can be clearly seen both on the points and solid lines. Similar to the QBER effects, afterpulses increase gains of the decoy1 and decoy2 pulses, especially decoy2, more than half of its gain comes from afterpulses. Finally, an average secure key rate of 306k bits/s is achieved with a failure probability of $10^{-10}$ \cite{yuan2013}. 

In summary, we develop a practical QKD system based on the asymmetric FSMI structure, which maintains the intrinsic stability of our previous FMI structure. Under channel disturbance, the interference fringe visibility of this practical system is concentrated around $99.35\%$ with a standard deviation of $0.12\%$ over a 24-h test. More importantly, the QKD system based on the FSMI structure shows high-speed support. Gigahertz QKD was implemented over a 50-km fiber channel with polarization scrambling, the QBERs and overall gains remain stable over 7 days without active polarization alignment. The results demonstrate that this intrinsically stable and high-speed QKD implementation is practical against channel disturbance, and is particularly suitable for deployment in field environments.

 This work was supported by the National Natural Science Foundation of China (Grant Nos. 61622506, 61575183, 61627820, and 61475148), the National Key Research And Development Program of China (Grant No. 2016YFA0302600), and the Strategic Priority Research Program (B) of the Chinese Academy of Sciences (Grant Nos. XDB01030100 and XDB01030300).

\end{document}